\documentclass[aps, amsfonts, longbibliography, reprint]{revtex4-1}
\usepackage{graphicx}
\usepackage{amsmath}
\begin{document}
\newcommand{\appref}[1]{(see also App.~\ref{suppl:#1})}
\newcommand{\figref}[1]{Fig.~\ref{fig:#1}}
\newcommand{\Figref}[1]{Figure \ref{fig:#1}}
\newcommand{\secref}[1]{Sec.~\ref{sec:#1}}
\newcommand{\mysection}[1]{\textit{#1.---}}
\newcommand{\deffig}[4]{%
  \begin{figure}[htb]
    \includegraphics[width=#2\textwidth]{#3}
    \caption{#4}
    \label{fig:#1}
  \end{figure}
}
\newcommand{\Renyi}{R\'{e}nyi }
\title{
  Entropy Governed by the Absorbing State of Directed Percolation
}
\author{Kenji Harada}
\affiliation{Graduate School of Informatics, Kyoto University, Kyoto 606-8501, Japan}
\author{Naoki Kawashima}
\affiliation{Institute for Solid State Physics, University of Tokyo, Kashiwa, Chiba 277-8581, Japan}
\begin{abstract}
  We investigate the informational aspect of (1+1)-dimensional
  directed percolation, a canonical model of a nonequilibrium
  continuous transition to a phase dominated by a single special state
  called the ``absorbing'' state. Using a tensor network scheme, we
  numerically calculate the time evolution of state probability
  distribution of directed percolation. We find a universal relaxation
  of \Renyi entropy at the absorbing phase transition point as well as
  a new singularity in the active phase, slightly but distinctly away
  from the absorbing transition point. At the new singular point, the
  second-order \Renyi entropy has a clear cusp. There we also detect a
  singular behavior of ``entanglement entropy,'' defined by regarding
  the probability distribution as a wave function. The entanglement
  entropy vanishes below the singular point and stays finite above. We
  confirm that the absorbing state, though its occurrence is
  exponentially rare in the active phase, is responsible for these
  phenomena. This interpretation provides us with a unified
  understanding of time evolution of the \Renyi entropy at the
  critical point as well as in the active phase.
\end{abstract}
\maketitle
%
%
\mysection{Introduction}
We can define the phase of a macroscopic system out of equilibrium
based on physical order parameters of a nonequilibrium steady-state
distribution. The phase transition is confirmed theoretically and
experimentally in many studies\cite{Derrida:1993bt, Sasamoto:1999fl,
  Domany:1984fh, Takeuchi:2007co, Takeuchi:2009ia, Sipos:2011bq,
  Sano:2016kh, Derrida:2007hr, Henkel:2008vv}.  However, we have
little understanding of the informational aspect of a macroscopic
nonequilibrium system.  Recently, Wood {\it et al.}
\cite{Wood:2017gt} analytically calculated the statistical \Renyi
entropy of the asymmetric exclusion process \cite{Derrida:1993bt,
  Sasamoto:1999fl, Derrida:2007hr}. The phase boundary defined by the
behavior of the \Renyi entropy agrees with the conventional phase
boundary of the asymmetric exclusion process for three nonequilibrium
phases. It shows the potential power of informational
quantities. However, computation of the entropy is technically
difficult; no method for directly measuring entropy is known for the
Monte Carlo simulation of a nonequilibrium system. Therefore, the
knowledge about the informational aspect of a nonequilibrium system is
limited than that of an equilibrium system\cite{Barnett:2013ix,
  Iaconis:2013, Stephan:2014kt}.

In this study, overcoming the technical difficulty by the tensor
network method, we will focus on the informational aspect of a
$(1+1)$-dimensional directed percolation (DP) through the \Renyi
entropy and the entanglement entropy. While the \Renyi entropy is
discussed frequently in conventional statistics, the introduction of
the entanglement entropy to the analysis of the classical model is
somehow unconventional. Its utility in the context of the classical
models was recently discussed \cite{Banuls:2019vi, Helms:2019vg}. If
we regard a percolating direction of objects as a time direction, the
$(d+1)$-dimensional DP corresponds to the $d$-dimensional
reaction-diffusion process. There are generally two phases in a
reaction-diffusion process, the active phase with finite density of
objects and the inactive phase with zero density in the long-time
limit. As shown in many numerical studies for DP(see a review
\cite{Henkel:2008vv}), the transition between the active phase and the
inactive phase is critical, and the concept of the universality is
also extended as the DP universality class. Because of the extreme
simplicity, the universality class is expected to be
ubiquitous\cite{Takeuchi:2007co, Takeuchi:2009ia, Sipos:2011bq,
  Vazquez:2011gn, Sano:2016kh, Filho:2018gm, Deng:2018ti}. In the
following, after we first briefly introduce the model of DP and the
numerical method for the \Renyi entropy and the entanglement entropy,
we will report the results of both entropies as well as their
interpretations.

%
%
%
\mysection{Directed percolation}
A (1+1)-dimensional DP is defined on a square lattice rotated by
45$^\circ$. A site can be either active or inactive. An active state
can percolate to the nearest-neighbor sites with a probability
$p$. However, the direction of percolation is limited. If it is
downward, there are two nearest-neighbor sites from which a site may
be percolated. We can regard the row of sites as a one-dimensional
system at a time. Then, the DP describes a reaction-diffusion process
in a one-dimensional system. This DP is called a bond DP, and it is a
special case of Domany-Kinzel (DK) automaton\cite{Domany:1984fh}. In
the DK automaton, a probability of an active state depends on the
number of active states $n$ in the nearest-neighbor sites at the
previous time as $P[n]$. If there is no active nearest neighbor just
before the current time, the probability of having an active state is
zero, i.e., $P[0]=0$. The bond DP is defined as $P[1]=p,
P[2]=p(2-p)$. There is a special state in a DK automaton in which
there is no active site. It is called the absorbing state because a
system cannot escape from it. If $P[1]$ and $P[2]$ are small, the
state of the system eventually arrives at the absorbing state after a
long time. Then, the evolution of the automaton is stopped. If $P[1]$
and $P[2]$ are large enough, the steady-state distribution with other
states can exist in the thermodynamic limit. Thus, there are an active
phase and an inactive phase in a DK automaton, and the nonequilibrium
phase transition between them is critical. It is called an absorbing
phase transition. The behavior at the critical point of DK automatons
is universal. It is called the DP universality class.

%
%
\mysection{Numerical method for \Renyi and entanglement entropies}
A state probability distribution, $P(S)$, can be mapped to a
normalized ``{wave function}'' \cite{Doi:1976dy} of which amplitude is
proportional to a state probability distribution as
$\vert \psi(t) \rangle = {\exp[H_2(t)/2]}{\sum_{S} P(S) \ \vert S
  \rangle}$.  Here, $H_2$ is the second-order \Renyi
entropy\cite{Renyi:1961ty}. It is a generalization of the Shannon
entropy and is defined as $H_q = -(q-1)^{-1}\log\sum_S [P(S)]^q$,
where $q$ is called the order. The Shannon entropy is recovered when
$q \to 1$.

Recently, various new techniques have been developed in the
calculation of a wave function. The most promising approach is the use
of a tensor network representation which is a composite tensor defined
by tensor contractions. Here, we use a one-dimensional tensor network,
i.e., matrix product states (MPS), to represent a wave function of a
one-dimensional system. Since the transfer matrix in a master equation
is an operator to the wave function which is written as a tensor
network\appref{TM_DK}, we can calculate the time step evolution of
a wave function as a tensor contraction between
them\cite{Johnson:2010gg, Johnson:2015eq, Hotta:2016bc}. Holding a
canonical form of MPS\cite{Shi:2006hz}, we can efficiently calculate
an approximated MPS at the next time step by the time-evolving block
decimation method\cite{Vidal:2003gb, Vidal:2004jc}\appref{CF}. We
can control the precision of MPS representation by a bond dimension of
the virtual index between two tensors in MPS\appref{Density}.
Since the second-order \Renyi entropy appears in a normalization
factor of a wave function, we can efficiently calculate it from a
tensor contraction of two MPSs.

Through the above mapping between a distribution and a wave function,
we can formally introduce the concept of entanglement of a quantum
state into a state probability distribution. The entanglement entropy
of a quantum state is the Shannon entropy of a quantum subsystem as
$H_E(A) = -\mathrm{Tr} \ \rho_A \log \rho_A$, where $\rho_A$ is a
reduced density operator of a subsystem $A$. The entanglement entropy
quantifies a quantum correlation between a subsystem $A$ and its
complement $\bar{A}$. In general, we can define the Schmidt
decomposition of a quantum state as
$\vert \psi \rangle = \sum_k e^{-(1/2)\xi_k} \vert \psi_A^k\rangle
\otimes \vert \psi_{\bar{A}}^k\rangle$, where $\vert \psi_A^k\rangle$
and $\vert \psi_{\bar{A}}^k\rangle$ are orthogonal states in
subsystem $A$ and its complement $\bar{A}$, respectively. Then, the
entanglement entropy is rewritten as
$H_E(A) = \sum_k \xi_k\exp(-\xi_k)$ and $\{\xi_k\}$ is called an
entanglement spectrum. Since the entanglement spectrum is a key
component in a canonical form of MPS, we can efficiently calculate it
from the canonical form of MPS.

%
%
%
\mysection{Critical relaxation of \Renyi entropy}
The density of active sites is an order parameter of DP. If we start
from a homogeneously full active initial state, the density of active
sites shows a power-law decay at the critical point.
For example, critical points of the bond DP and the site DP are at
$p_c = 0.644700185(5)$ and $0.70548515(20)$, respectively
\cite{Jensen:2004dm}.
The decay exponent is universal in the DP universality
class. \Figref{relax} shows the critical relaxation of the density and
the \Renyi entropy per site at the critical points of the bond DP and
the site DP of $8192$ sites with open boundary condition. The tensor
network with the bond dimension 250 was used for the calculation.
Here, to reduce the boundary effect, the density is measured at the
central site, whereas the \Renyi entropy is defined on the central
half of the system. The site DP is a special case in DK automaton as
$P[0]=0, P[1]=P[2]=p$, and it also belongs to the DP universality
class. Thus, the decay exponents for both DPs should be equal. Our
estimates of the decay exponent based on \figref{relax},
$\delta_{\textrm{bond}} = 0.1593(1)$ for the bond DP and
$\delta_{\textrm{site}} = 0.158(1)$ for the site DP. They are
consistent with a previously estimated value $\delta = 0.159464$
\cite{Jensen:1999jl} within the error. Thus, we confirm the
reliability of MPS calculation for criticality. The \Renyi entropy per
site is also an order parameter of the absorbing phase transition
because it takes a finite value only in the active phase. As shown in
\figref{relax}, \Renyi entropy at the critical point also shows a
power-law decay. Since the decay exponents in \figref{relax} are
$0.632(5)$ and $0.625(7)$ for bond DP and site DP, respectively, we
confirm the universal relaxation in the \Renyi entropy.
As we see below, this universal relaxation can be expected as
\begin{equation}
  \label{eq:h2c}
  h_2(t) = H_2(t)/N \sim \xi_\bot(t)^{-1} \sim t^{-1/z},
\end{equation}
where $\xi_\bot$ is the spatial correlation length. Indeed, the value of
decay exponent in \figref{relax} is consistent with the reciprocal of
the dynamical critical exponent,
$1/z=0.632613(6)$\cite{Jensen:1999jl}.

\deffig{relax}{0.5}{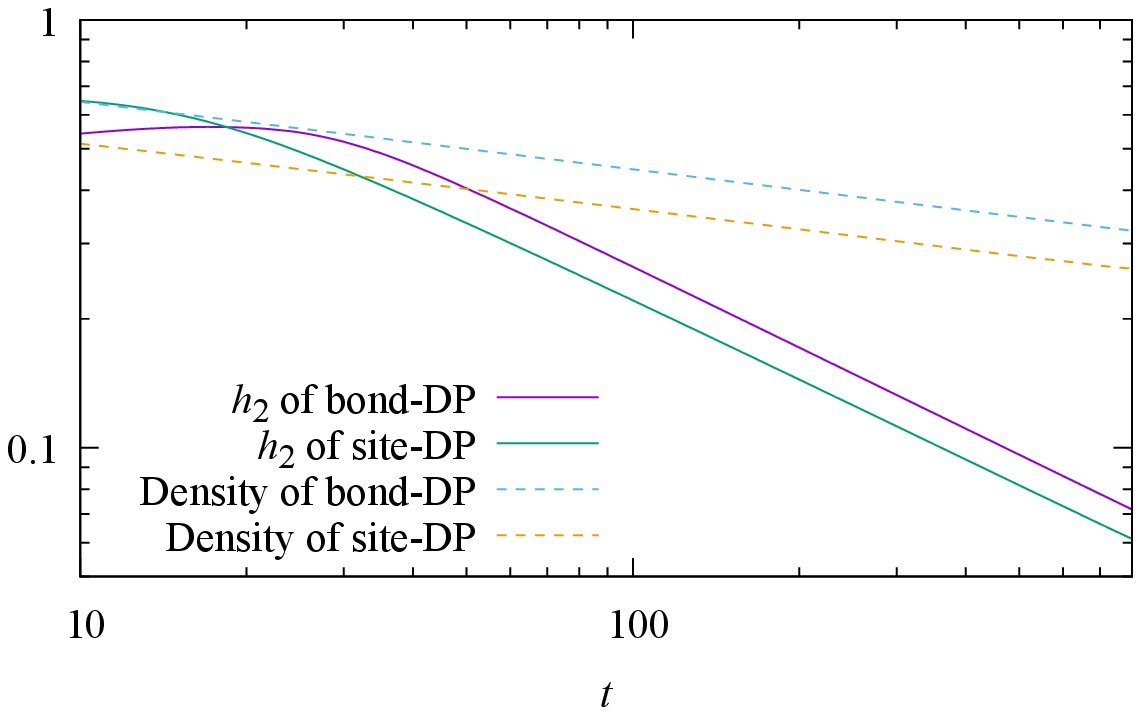}{%
  Time evolution of the second-order \Renyi entropy per site, $h_2$,
  (solid) and the density of active sites (dashed) at critical points
  of bond DP and site DP.
}
\mysection{\Renyi entropy and entanglement entropy in active phase}
If the initial probability distribution consists of a fully active
state, the \Renyi entropy per site starts from zero. In the early time
region, it rapidly grows. After that, it quickly converges to a finite
value in the active phase.  \Figref{re_conv} shows the \Renyi entropy
per site extrapolated to the thermodynamic limit by a $1/N$ quadratic
fitting for a system size $N \le 4096$\appref{Renyi}. As shown in
the inset, there is a cusp at $p_2^* = 0.6785(5)$ in the active phase.

\deffig{re_conv}{0.5}{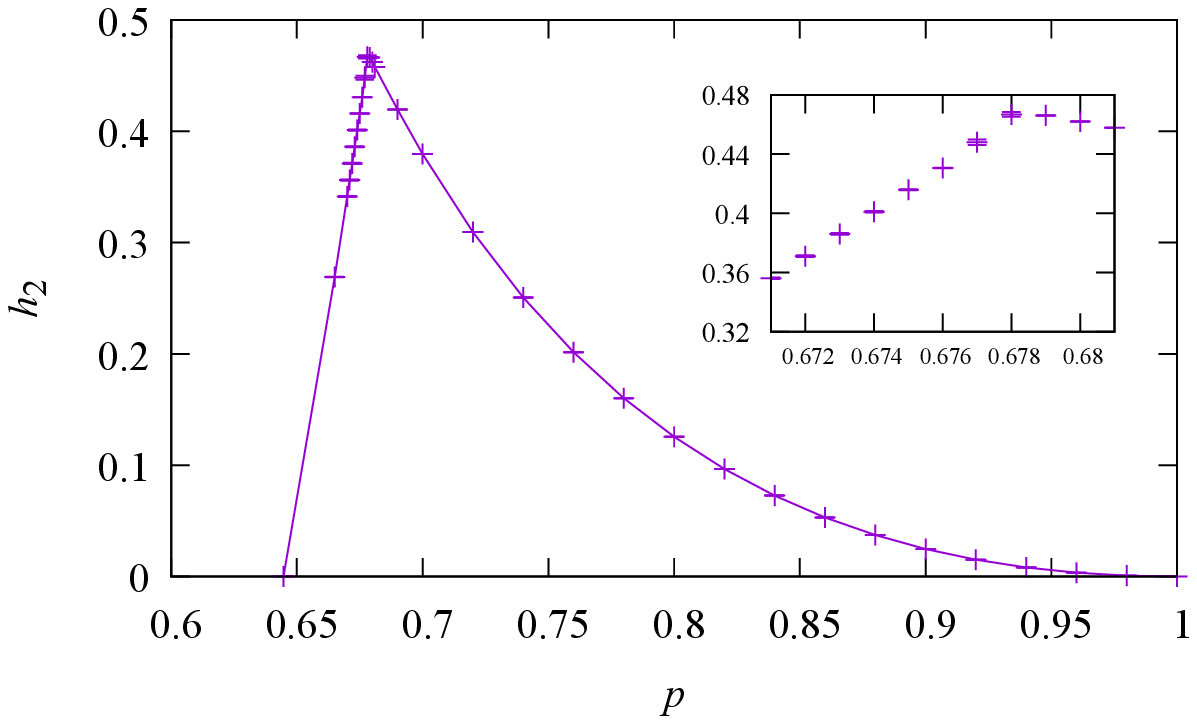}{%
  The second-order \Renyi entropy per site $h_2$ of the steady-state
  distribution of the bond DP in the thermodynamic limit. There is a
  clear cusp slightly above the bond-DP threshold,
  $p_c = 0.644700185(5)$ \cite{Jensen:2004dm}. The inset is the
  enlarged view near $p_2^* = 0.6785(5)$.
}

We calculate the time evolution of entanglement entropy between the
left and the right halves of the system from the fully active initial
state.  \Figref{ee_es} shows entanglement entropy from the system size
128 to 8192. The bond dimension in MPS is 50\footnote{Results in
  \figref{ee_es} do not change when the bond dimension is larger than
  50.}. Solid and dashed lines represent entanglement entropy of
$p=0.678$ and $0.679$, respectively. The entanglement entropy has a
peak for both cases. The peak time of $p=0.678$ almost converges for
more than 4096 sites. However, the peak time of $p=0.679$ diverges
with the system size. Solid and dashed lines in the inset represent
the levels in the entanglement spectrum from $\xi_0$ to $\xi_2$ for
$p=0.678$ and $0.679$ with 512 sites, respectively. The crossover time
at which two levels $\xi_0$ and $\xi_1$ cross agrees with the peak
time of entanglement entropy. \Figref{tau0_gap} shows the crossover
time, $\tau_0$, from $p=0.676$ to $0.681$ for less than 8192
sites. When we increase the system size, $\tau_0$ diverges for
$p \ge 0.6786$. However, $\tau_0$ converges or its increase becomes
weaker for $p \le 0.6785$. In the thermodynamic limit, the
entanglement entropy has a finite peak time for
$p < p_E^* = 0.67855(5)$, but no peak for $p > p_E^*$. The asymptotic
behavior of entanglement entropy also changes at $p_E^*$. In a long
time region, the entanglement entropy eventually decreases and
converges to zero for $p < p_E^*$. On the other hand, it converges to
a finite value for $p > p_E^*$.  As shown in the inset, the asymptotic
behavior of a gap $\Delta_0$ between $\xi_0(\tau_0)$ and
$\xi_1(\tau_0)$ also changes at $p_E^*$. The gap is open for
$p < p_E^*$, but it is closed for $p > p_E^*$. The position of a
singularity of entanglement entropy $p_E^*$ is equal to the position
of a singularity of the second-order \Renyi entropy $p_2^*$ within the
error. As we see below, these singularities can be explained by the
effect of the extremely rare absorbing state.

\deffig{ee_es}{0.5}{{ee_es_0.678_0.679_512}.eps}{%
  Time evolution of entanglement entropy $H_E$ between the left and
  the right halves of the system from system size 128 to 8192. Solid
  and dashed lines represent $H_E$ at $p=0.678$ and $0.679$,
  respectively. Solid and dashed lines in the inset represent singular
  values $\xi_0, \xi_1$ and $\xi_2$ at $p=0.678$ and $0.679$ for $512$
  sites, respectively.  }
\deffig{tau0_gap}{0.5}{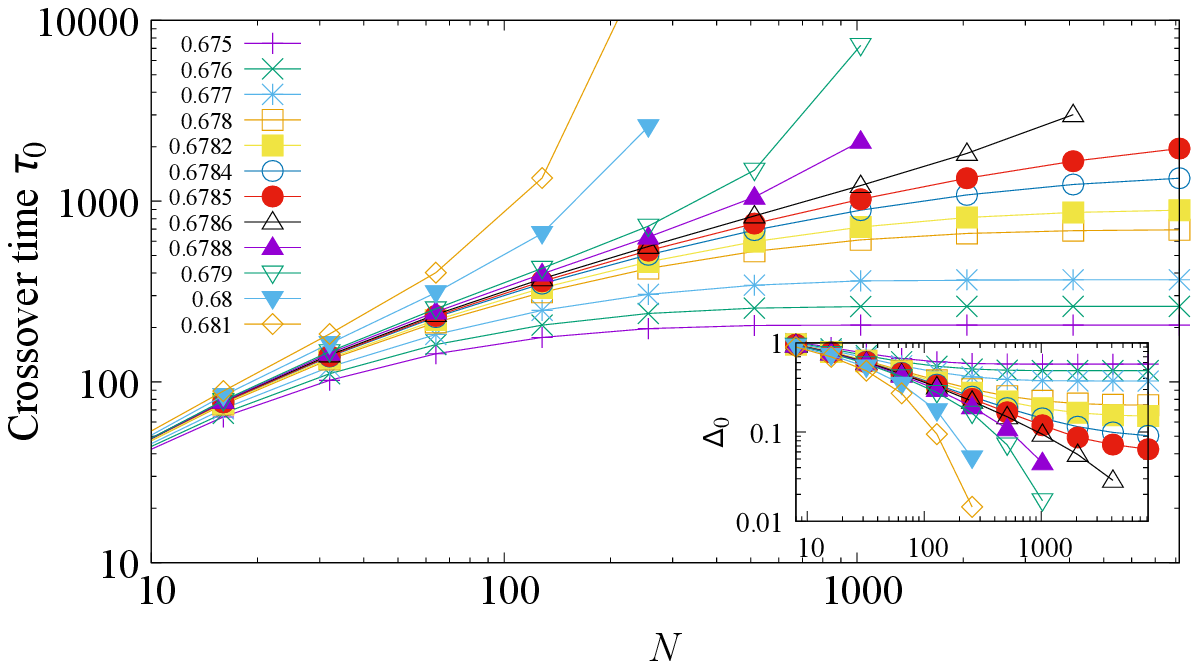}{%
  Crossover time $\tau_0$ between two levels $\xi_0$
  and $\xi_1$ from $p=0.676$ to $0.681$. $N$ denotes a system size.
  The inset shows a gap $\Delta_0$ between $\xi_0(\tau_0)$ and
  $\xi_1(\tau_0)$.
}
%

%
%
\mysection{Effect of absorbing state}
The probability of the absorbing state monotonically increases,
because the system cannot escape from it. Therefore, the absorbing
state is the only steady state for any finite system. To define a
nontrivial steady state, we must consider an infinite system. We then
consider a marginal distribution for a subsystem that consists of
sequential $N$ sites. In the long time limit, it safely converges to a
steady-marginal distribution $P_N(S)$. From our calculation, we
observe that this steady-state marginal distribution consists of two
components, the delta-function at the zero-density state and the
continuous component around some nonzero density. Namely,
\begin{equation}
  \label{eq:pn}
  P_N(S) = \delta_{S=0}P_N(0) +  \delta_{S\ne 0} (1-P_N(0)) Q_N(S).
\end{equation}
The delta-function component can be interpreted as a
``{quasiabsorbing}'' state, which is a consequence of the fact
that the zero-density state plays a special role even in the marginal
distribution because of the absence of the nucleation, i.e., active
sites occur only by the influence from the outside of the boundary.
Here, since the occurrence of the absorbing state is exponentially
rare in the active phase, we assume an asymptotic form of the
probability of quasiabsorbing state as
\begin{equation}
  \label{eq:p0}
  P_N(0) \sim \exp\left[-\sigma_0({N}/{\xi_\bot})\right],
\end{equation}
where $\exp(-\sigma_0)$ is the probability of having the zero-density
state in a correlated region where $\xi_{\bot}$ is the region's
size. In other words, $\sigma_0$ is the renormalized chemical potential
(of holes).

In the active phase, the \Renyi entropy per site of a continuous-part $Q_N$
converges to $h_q' \equiv -(q-1)^{-1}\log\sum_{S}[Q_N(S)^q]/N$. Using
\eqref{eq:p0}, \Renyi entropy per site for a steady-marginal distribution
in the large $N$ limit is written as
\begin{equation}
  \label{eq:renyi}
  h_q = \min\left[\left\{{q}/{(q-1)}\right\} \left({\sigma_0}/{\xi_\bot}\right), \ h_q'\right].
\end{equation}
The singularity at $p_q^*$ can be understood as the crossover from the
first-term dominant region to the second term dominant one, so that
$\{q/(q-1)\}\{\sigma_0(p_q^*)/\xi_\bot(p_q^*)\} =
h_q'(p_q^*)$\footnote{Based on \eqref{eq:renyi}, we can expect the
  singular behavior of the \Renyi entropy near $p_c$ with the exponent
  $\nu_\bot$. The value of the exponent estimated based on all points
  of \figref{re_conv} below $p_2^*$ is $1.093(5)$. Our estimation of
  the exponent is consistent with the value of
  $\nu_\bot = 1.096854(4)$\cite{Jensen:1999jl}.}.

The entanglement entropy is the Shannon entropy for the singular value
part of the amplitude matrix $\mathbb{P}$ as
$\mathbb{P}_{S_L, S_R} = \langle S_L, S_R\vert \psi \rangle$ where
$S_L$ and $S_R$ denote states of left and right half systems.  Based
on \eqref{eq:pn}, $\mathbb{P} = \mathbb{P}^0 + \mathbb{Q}$, where
${\mathbb{P}^0}_{S_L, S_R} = \exp(Nh_2/2) P_N(0)\delta_{S_L=S_R=0}$.
From \eqref{eq:renyi}, the Frobenius norm\footnote{The Frobenius norm
  of matrix $\mathbb{A}$ is
  $\sqrt{\textrm{Tr} \mathbb{A}^{\dagger}\mathbb{A}}$. It is the
  square root of the summation of squared singular values.}
$\vert \mathbb{P}^0 \vert \gg \vert \mathbb{Q}\vert$ for $p < p_2^*$,
and vice versa in the large $N$ limit. The quasiabsorbing state is a
direct product state. Therefore, the entanglement entropy is zero for
$p < p_2^*$ and may be finite for $p > p_2^*$. This result is
consistent with the common singular point for the entanglement entropy
and the second-order \Renyi entropy in the bond DP.

\deffig{comp_h2_p0}{0.5}{{comp_h2_p0}.eps}{%
  Comparison between the second-order \Renyi entropy per site $h_2$
  (solid lines) and the absorbing-state entropy, $-2\log[P_N(0)]/N$
  (symbols).
}

So far, we have been discussing singularity of the entropies due to
the crossover from the ``{absorbing-state dominating}'' regime
to the ``{continuous-part dominating}'' regime.  While the
transitions discussed above are fictitious, the ``\textit{real}''
critical phenomena can also be understood by this crossover.  To see
this, we extend \eqref{eq:p0} to the probability $P_N(0,t)$ allowing
the time dependence of the correlation length as
\begin{equation}
  \label{eq:dynamical_p0}
  P_N(0,t) \sim \exp[-\sigma_0N/\xi_{\bot}(t)].  
\end{equation}
As shown in \figref{comp_h2_p0}, the absorbing-state entropy
$-2\log P_N(0,t)/N$, equal to the first term of \eqref{eq:renyi}, is a
decreasing function of time, because $\xi_{\bot}(t)$ monotonically
increases. For $p = 0.68 > p_2^* = 0.6785(5)$, it never completely
catches up with $h_2(t)$. When $p_c < p = 0.67 < p_2^*$, it catches up
with it at a certain point of time and once it does it dominates the
\Renyi entropy; $h_2$ and the absorbing-state entropy agree with each
other afterward. These are the behaviors that are naturally expected
from \eqref{eq:renyi} and \eqref{eq:dynamical_p0}. However, this
behavior remains essentially the same even at the critical point
$p=p_c$, with the only difference being that the asymptotic curve is
power-law decay in \eqref{eq:h2c}. As shown in \figref{comp_h2_p0},
the agreement between the slope of the asymptotic behavior and $z$
confirms the validity of the interpretation with the simple assumption
\eqref{eq:dynamical_p0}.

%
%
\mysection{Conclusion}
We focus on the informational aspect of the (1+1)-dimensional DP
through a statistical \Renyi entropy and an entanglement entropy. We
numerically find a universal relaxation with a slope $1/z$ at the
critical point $p_c$, and a cusp at $p_2^*=0.6785(5)$ in the active
phase of the bond DP for the second-order \Renyi entropy. As we
increase $p$ passing $p^*_2$, the dynamical behavior of the
entanglement entropy also changes. The entanglement entropy
asymptotically goes to zero below $p^*_2$ and converges to a finite
value above $p^*_2$. Based on the assumption \eqref{eq:dynamical_p0}
for the asymptotic form of the probability of quasiabsorbing state, we
can understand all of their behavior as the domination of the
absorbing state of which the occurrence is exponentially rare in the
active phase. Since the existence of the absorbing state is the unique
character of DP, the universal relaxation and the cusp in the \Renyi
entropy are hallmarks of DP\appref{site_Renyi}. Our scenario can
be extended to the higher-dimensional DP and the DP-like models.

In this study, we show an application of a tensor network scheme to DP
as a canonical model of a nonequilibrium continuous transition. The
tensor network scheme gives us detailed information of the state
probability distribution as the \Renyi entropy. We hope for the
further use of tensor networks to understand the characterization of
nonequilibrium systems from the informational point of view.

\mysection{Acknowledgment}
K.~H. appreciates fruitful comments from T.~Okubo and K.~Tamai, and
K.H. acknowledges many discussions with R.~Sato and K.~Nataochi about
the numerical schemes in the early stage of this study. This work was
supported by JSPS KAKENHI Grant No. 17K05576, and by MEXT as
Exploratory Challenge on Post-K computer (Frontiers of Basic Science:
Challenging the Limits). The computation in this work has been done
using the facilities of the Supercomputer Center, the Institute for
Solid State Physics, the University of Tokyo and the facilities of the
Supercomputer Center at Kyoto University.

\appendix
\renewcommand{\thesection}{\arabic{section}}
\section{Exact tensor network representation of a transfer matrix of
  the Domany-Kinzel automaton}
\label{suppl:TM_DK}

The Domany-Kinzel (DK) automaton\cite{Domany:1984fh} is defined on a
square lattice rotated 45 degrees in \figref{DP_suppl}(a). A state on
a site can be either active or not. An active state can propagate to
the nearest neighbor sites at the next time. A probability of an
active state depends on the number of active states, $n$, in the
nearest neighboring sites at the previous time as $P[n]$. Always,
$P[0]=0$.  If the rotated square lattice of the DK automaton is
transformed to a regular lattice from \figref{DP_suppl}(a) to
\figref{DP_suppl}(b), the DK automaton is a stochastic process on a
chain.
\deffig{DP_suppl}{0.45}{{DP_suppl}.eps}{ (a) A DK automaton on a square
  lattice rotated 45 degrees. Filled and open circles represent active
  and inactive sites, respectively. The number of sites is four. There
  are two open boundaries. A dashed bond between circles represents
  the dependence of a state at the next time.  (b) A DK automaton on a
  regular square lattice which is equal to (a). The (yellow) thick
  line corresponds to the (yellow) thick line in (a).  }

Since an active state in the DK automaton is locally propagated, the
transfer matrix of DK automaton can be exactly rewritten as a tensor
network as shown in \figref{transfer_DBP_new}. This is a transfer
matrix of a two time-step evolution for four sites with two open
boundaries which corresponds to \figref{DP_suppl}(a).  The tensor $C$,
$V$, and $W$ are defined as
\begin{align}
  C_{ijk} & =
            \begin{cases}
              1 & \mbox{if $i=j=k$.}\\
              0 & \mbox{otherwise.}
            \end{cases}\\
  V_{ij} &= (1-j)(1-P[i])+j P[i]\\
          &= (1-j) + (2j-1)P[i]\\
  W_{lmn} &= (1-n)(1 - P[l+m]) + n P[l+m]\\
          &= (1-n) + (2n-1)P[l+m].
\end{align}
Here each index is a binary variable, 0 or 1. The composite tensor of
$C$ and $W$ enclosed by a dashed line is a two-site operator. At an
edge of a chain, it also includes the tensor $V$.  Therefore, there
are four different composite tensors. The transfer matrix of a two
time-step evolution is a cascade of these composite tensors (two-site
operators) from left to right and the opposite.
\deffig{transfer_DBP_new}{0.35}{{transfer_DBP_new}.eps}{Tensor network
  representation of a transfer matrix of a two time-step evolution for
  the DK automaton of four sites with two open boundaries.}
\section{Canonical form of a matrix product state}
\label{suppl:CF}

A canonical form of a matrix product state(MPS)\cite{Shi:2006hz}
consists of three types of tensors as $L$, $R$, and $\Lambda$ shown in
\figref{CF}(a).  $L$ and $R$ are isometries as follows:
\begin{equation}
 \sum_{ij}L_{ijk} L^*_{ijk'} = \delta_{kk'}, \quad \sum_{mn}R_{lmn} R^*_{l'mn} = \delta_{ll'}.
\end{equation}
$\Lambda$ is a positive diagonal matrix. Based on a singular value
decomposition(SVD)\cite{Shi:2006hz}, we can transform an MPS to a
canonical form. If we apply a two-site operator to a
$\Lambda$-neighboring site in a canonical form(see \figref{CF}(b)), we
can apply an SVD to a composite tensor from $R'$, $R''$, $\Lambda$,
and a two-site operator to obtain a new approximated MPS. This method
is called the time-evolving block decimation
method\cite{Vidal:2003gb, Vidal:2004jc}. Then, the new MPS again takes
a new canonical form in which the position of $\Lambda$ is shifted to
right.  This SVD approximation means a global minimization of the norm
as $||\ \vert \psi' \rangle - T_{i,i+1} \vert \psi \rangle\ ||^2$,
where $T_{i,i+1}$ is a two-site operator.  The exact tensor network
representation of a transfer matrix of the DK automaton is a sequence
of two-site operators as shown in \figref{transfer_DBP_new}. Using a
canonical form of MPS, we can keep a total accuracy in the time
evolution of the DK automaton. Unlike in \cite{Johnson:2010gg,
  Johnson:2015eq, Hotta:2016bc}, if we do not use a canonical form, it
is hard to calculate the long time behavior of DK automaton with high
accuracy.
\deffig{CF}{0.45}{{canonical_form}.eps}{(a) A canonical form of
  MPS. (b) A new canonical form of MPS operated by a two-site
  operator.}

\section{Time evolution of a density in the bond DP}
\label{suppl:Density}

\deffig{Density_DBP}{0.5}{{density_time}.eps}{The density in bond DPs
  from $p=0.6$ to $p=0.7$. Solid lines represent the average density
  of two sites at the center of a system. The bond dimension in a
  canonical form is $D=120$. For all cases, the initial state is a
  full active state. The system size is $N=512$ except for $N=256$ at
  $p=0.7$. Symbols represent Monte Carlo results for the same system
  size.}
In our numerical simulation, we use a canonical form of MPS which
approximates a state probability distribution. The bond dimension of a
canonical form should be large enough for a high
precision. \figref{Density_DBP} shows the time evolution of a density
in the bond DP by using a canonical form with a bond dimension
$D=120$. Here, in order to reduce the effect of open boundaries, the
density is defined as the average density of two sites at the center
of a system. The system size is $N=512$ except for $N=256$ at
$p=0.7$. Results of the canonical form are consistent with those of
Monte Carlo simulations for the same system size. There is no visible
dependence on the finite bond dimension and the finite system size for
these results.

\section{\Renyi entropy of the steady-state distribution of the
  bond DP in the thermodynamic limit}
\label{suppl:Renyi}

In order to estimate a \Renyi entropy, we need to check the dependence
on the system size and the bond dimension $D$ of
MPS. \figref{Chi_dependence_Renyi} shows the time evolution of the
\Renyi entropy of the bond DP by $D=50$, and $100$ for $N=4096$.
There is no visible difference between $D=50$ and $D=100$.

The \Renyi entropy quickly converges to the value of the steady-state
distribution before $t=2000$. \figref{Convergence_Renyi} shows the
values of \Renyi entropy near $p_2^* = 0.6785(5)$ at $t=2000$. There
is the strong finite-size effect of a system size $N$. Using $1/N$
quadratic fitting curves, we estimated the value of the \Renyi entropy
in the thermodynamic limit. \figref{Thermodynamic_limit_Renyi} shows
the \Renyi entropy in the thermodynamic limit. The error shown in
\figref{Thermodynamic_limit_Renyi} (the inset of Fig.2 of the main
text) is the extrapolation error with respect to the system size. The
error due to the finiteness of $D$ is much smaller. As shown in
\figref{Thermodynamic_limit_Renyi}, there is a cusp slightly above the
bond-DP threshold, $p_c = 0.644700185(5)$ \cite{Jensen:2004dm}. The
position of the cusp is between $p = 0.678$ and $0.679$. Therefore, we
read the error in our estimate of $p_2^*$ is smaller than $0.0005$ as
$p_2^* = 0.6785(5)$.

\deffig{Chi_dependence_Renyi}{0.5}{{re_conv_check_chi}.eps}{Time
  evolution of the \Renyi entropy per site, $h_2$, of the bond-DP
  steady-state distribution. $D$ is a bond dimension of MPS. The
  system size is $4096$. Inset: the \Renyi entropy per site up to
  $t=2000$. }
\deffig{Convergence_Renyi}{0.5}{{re_conv_check_fs}.eps}{Finite-size
  effect of the \Renyi entropy per site, $h_2$, of the bond-DP
  steady-state distribution.  $N$ is system size. Symbols denote the
  \Renyi entropy per site of a finite size system. Solid lines denote
  $1/N$ quadratic fitting curves for $N \le 4096$.}
\deffig{Thermodynamic_limit_Renyi}{0.5}{{re_conv_inset}.eps}{\Renyi
  entropy per site, $h_2$, of the bond-DP steady-state distribution in
  the thermodynamic limit.  There is a cusp slightly above the bond-DP
  threshold, $p_c = 0.644700185(5)$ \cite{Jensen:2004dm}. }
\deffig{Convergence_Renyi_sdp}{0.5}{{re_conv_sdp}.eps}{ \Renyi entropy
  per site, $h_2$, of the site-DP steady-state distribution in the
  thermodynamic limit. There is a cusp slightly above the site-DP
  threshold, $p_c = 0.70548515(20)$ \cite{Jensen:2004dm}.  }

\section{\Renyi entropy of the steady-state
  distribution of the site DP in the thermodynamic limit}
\label{suppl:site_Renyi}

Using the same procedure in the bond-DP case, we estimated the \Renyi
entropy of the site DP in the thermodynamic limit from the data of
$N \le 1024$. As shown in \figref{Convergence_Renyi_sdp}, there is a
similar cusp in the \Renyi entropy of the site DP in the thermodynamic
limit. The position of cusp is at $p_2^* = 0.756(1)$ slightly above
the site-DP threshold, $p_c = 0.70548515(20)$ \cite{Jensen:2004dm}.

\bibliography{dp}
\end{document}